\documentclass[conference]{IEEEtran}
\IEEEoverridecommandlockouts

\usepackage{cite}
\usepackage{amsmath,amssymb,amsfonts}
\usepackage{algorithmic}
\usepackage{graphicx}
\usepackage{textcomp}
\usepackage{xcolor}
\def\BibTeX{{\rm B\kern-.05em{\sc i\kern-.025em b}\kern-.08em
    T\kern-.1667em\lower.7ex\hbox{E}\kern-.125emX}}

\usepackage{subfigure}
\usepackage{tabularray}
\usepackage{multirow}
\usepackage[hidelinks,pdfencoding=auto]{hyperref}
\usepackage{enumitem}
\usepackage{stfloats}
\usepackage{tablefootnote}
\usepackage{placeins}
\usepackage{color,soul}
\usepackage{threeparttable}

\title{A Simultaneous ECG-PCG Acquisition System with Real-Time Burst-Adaptive Noise Cancellation}

\author{
Avishka Herath\textsuperscript{$\dagger$1*},
Malith Jayalath\textsuperscript{$\dagger$1}, 
Kumudu Kaushalya\textsuperscript{1},
Sanjana Kapukotuwa\textsuperscript{1},
Chathuni Wijegunawardena\textsuperscript{1}, \\
Pahan Mendis\textsuperscript{1}, 
Kithmin Wickremasinghe\textsuperscript{2},
Duminda Samarasinghe\textsuperscript{3},
Wageesha N. Manamperi\textsuperscript{1}, \\
and Chamira U. S. Edussooriya\textsuperscript{1*}

\thanks{\textsuperscript{$\dagger$}These authors contributed equally to this work.}
\thanks{\textsuperscript{1}Department of Electronic and Telecommunication Engineering, University of Moratuwa, Bandaranayake Mawatha, Moratuwa 10400, Sri Lanka.}
\thanks{\textsuperscript{2}Department of Electrical and Computer Engineering, University of British Columbia, 6200 University Blvd, Vancouver, BC V6T 1Z4, Canada.}
\thanks{\textsuperscript{3}Lady Ridgeway Hospital for Children, Dr Danister De Silva Mawatha, Colombo 00800, Sri Lanka.}
\thanks{\textsuperscript{*}Corresponding author e-mails: \href{mailto:avishkah@uom.lk}{avishkah@uom.lk}, \href{mailto:chamira@uom.lk}{chamira@uom.lk}\newline}
\thanks{This is the author’s version of the article that has been accepted and presented at the 48\textsuperscript{th} Annual International Conference of the IEEE Engineering in Medicine and Biology Society (EMBC 2026), Toronto, Canada.}
\thanks{\textcopyright~2026 IEEE. Personal use of this material is permitted. Permission from IEEE must be obtained for all other uses, 
in any current or future media, including reprinting/republishing this material for advertising or promotional purposes, creating new collective works, for resale or redistribution to servers or lists, or reuse of any copyrighted component of this work in other works.}
}

\begin{document}

\maketitle

\begin{abstract}
Cardiac auscultation is an essential clinical skill, requiring excellent hearing to distinguish subtle differences in timing and pitch of heart sounds. However, diagnosing solely from these sounds is often challenging due to interference from surrounding noise, and the information may be limited. Most of the existing solutions that adaptively cancel external noise are either non-real-time or computationally intensive, making them unsuitable for implementation in a portable system. 
This work proposes an end-to-end system with a real-time adaptive noise cancellation pipeline integrated into a device that simultaneously acquires electrocardiogram (ECG) and phonocardiogram (PCG) signals. We employ a burst-adaptive normalized least mean square algorithm that adjusts its adaptation in response to high-energy, non-stationary hospital noise. 
The algorithm's performance was initially assessed using datasets with artificially induced noise. Subsequently, the complete end-to-end system was validated using real-world hospital recordings captured with the dual-modality device. For ECG and PCG signals recorded from the device in noisy hospital settings, the proposed system achieved signal-to-noise ratio improvements of 30.32~dB and 37.01~dB, respectively. Furthermore, complexity analysis confirms the pipeline’s suitability for embedded implementation. These results demonstrate the system’s effectiveness in enabling reliable and accessible cardiac screening in noisy hospital environments typical of resource-constrained settings. 
\end{abstract}

\begin{IEEEkeywords}
Real-time Adaptive Noise Cancellation, Cardiac Auscultation, Phonocardiogram, Electrocardiogram
\end{IEEEkeywords}

\section{Introduction}

Cardiovascular diseases (CVDs) primarily arise from structural abnormalities of the heart, which may be congenital (present from birth) or acquired later in life. Congenital heart diseases affect approximately 9.41 per 1000 live births~\cite{A8}, causing more than 260,000 global deaths in 2017~\cite{A9}, and were estimated to globally account for over 200,000 deaths in children under five in 2021~\cite{xu2025global}. Meanwhile, CVDs in general caused an estimated 19.8 million deaths worldwide in 2022 (around 32\% globally), and three-quarters of CVD deaths take place in developing (low- and middle-income) countries~\cite{WHO_CVDs_2025}.

The stethoscope is traditionally considered the primary tool ubiquitously used for the initial assessment of cardiac conditions. However, considerable expertise is required for an accurate diagnosis using a stethoscope~\cite{A5}, and noise conditions of the surrounding clinical environments make this task further challenging~\cite{A18}.
Complementary to cardiac auscultation, the electrocardiogram (ECG) is an effective, non-invasive diagnostic tool that reflects the electrical activity of the heart.

Since the cyclic production of ECG and heart sounds (phonocardiogram, PCG) is biologically synchronized, the combination of these two modalities makes the localization of murmurs within the cardiac cycle as well as the identification of the first (S1) and the second (S2) heart sounds significantly more accurate~\cite{A3}. Diagnosing this using a traditional stethoscope requires considerable experience, as distinct heart murmurs, which are frequent indicators of cardiac abnormalities, can be easily missed in clinical diagnosis by a semi-trained healthcare professional~\cite{A5}. 
Furthermore, several studies have demonstrated that combined ECG-PCG data can be used to estimate clinically meaningful parameters that reflect the heart’s contractile function and pumping capacity~\cite{s20072033, paiva2009assessing, jimenez2021timing, zhang2025nmcse}. Recent research on machine learning techniques also shows that incorporating and/or coupling ECG and PCG modalities can significantly improve classification accuracy between normal and pathological cardiac conditions~\cite{9401093, li2022multi, zhang2024co}.

Due to the advancements in sensor technologies and signal processing techniques, digital stethoscopes have gained an edge over the traditional stethoscope. Today, widely used commercial digital stethoscopes can facilitate volume amplification, ambient noise reduction, and wireless transmission of sound signals between remote processing devices~\cite{littmann,web2,web3,web4}. However, many of these digital stethoscopes lack the functionality of synchronized ECG and PCG acquisition with wireless data transmission and real-time noise cancellation in hospital settings. Existing solutions that overcome these problems are unavailable in most developing countries, e.g., Eko Core 500\cite{web3}. 

In parallel, recent research has also explored wearable systems for integrated ECG-PCG monitoring~\cite{zang2025novel,A5,yoo2023wireless} and synchronous ECG-PCG auscultation platforms~\cite{lin2025portable,10569070}.
Despite these advancements, recording high-quality PCG signals remains challenging due to their susceptibility to ambient noise, especially in crowded or low-resource hospital environments. Such noise can distort the characteristic features of PCG signals and degrade diagnostic performance. 

Consequently, denoising the PCG signal is crucial before any further analysis. Prior studies have addressed PCG denoising through multistage adaptive least mean squares (LMS) filter architectures~\cite{pauline2022robust}, deep learning-based (DL) frameworks~\cite{10173517}, and optimization-driven adaptive filtering approaches~\cite{alla2025enhancing}. However, an open need remains for algorithms capable of adaptively denoising PCG signals in real-time while maintaining low computational complexity for embedded implementation. 

In this work, we propose a portable, end-to-end device for simultaneous ECG and PCG acquisition, featuring two integrated real-time signal processing pipelines. Unlike prior studies that focus mostly on offline denoising, our system processes both modalities in real-time with low computational complexity, making it ideal for embedded platforms. One pipeline performs ECG denoising, while the other utilizes real-time adaptive noise cancellation (ANC) for PCG signals via a novel burst-adaptive normalized LMS (BA-NLMS) filter. This filter is specifically designed to suppress the sudden, non-stationary noise common in hospital environments. The system was initially validated using synthetic data, followed by an experimental evaluation using real-world ECG-PCG recordings acquired in clinical settings. The results demonstrate that the proposed approach achieves average SNR improvements of 30.32~dB for ECG and 37.01~dB for PCG, confirming its robust noise suppression in real-world conditions and its suitability for deployment in accessible cardiovascular diagnostics.

\begin{figure*}[htbp]
    \centerline{\includegraphics[width=1\textwidth, trim=0pt 4pt 0pt 4pt, clip]{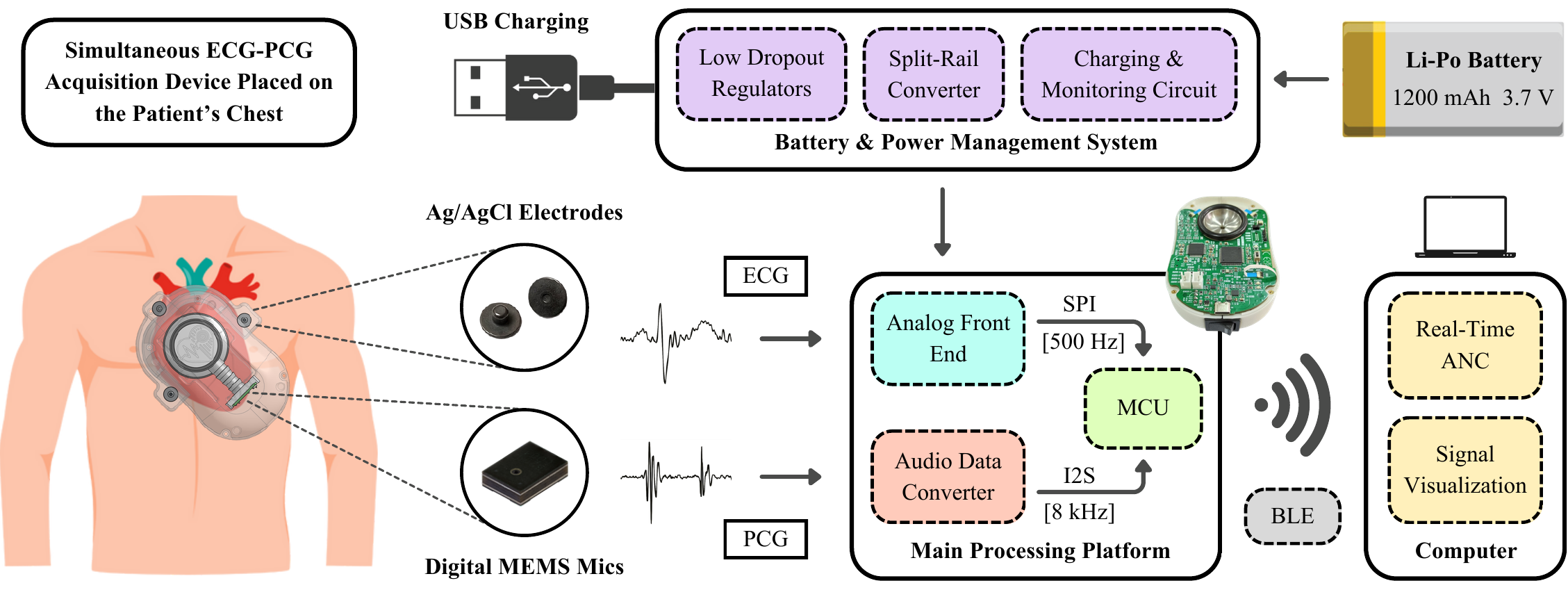}}
    \caption{\textbf{Block diagram of the overall system architecture.} Signals from dry Ag/AgCl electrodes (acquiring ECG) and digital microelectromechanical system (MEMS) microphones (acquiring PCG) are captured, processed by a microcontroller unit (MCU), and transmitted wirelessly to a computer for real-time ANC and signal visualization. The portable acquisition device is powered up using a rechargeable 3.7~V Li-Po battery.}
    \label{fig:block}
    \vspace{-0.4cm}
\end{figure*}

\section{System Overview}

The functional blocks of the overall system architecture for our dual-modality device used for comprehensive cardiac screening are depicted in Fig. \ref{fig:block}.
It integrates the acquisition, processing, transmission, and denoising of two biosignals, ECG and PCG, from the body in real-time.

\begin{figure}[!b]
    \vspace{-0.4cm}
    \centerline{\includegraphics[width=1\columnwidth]{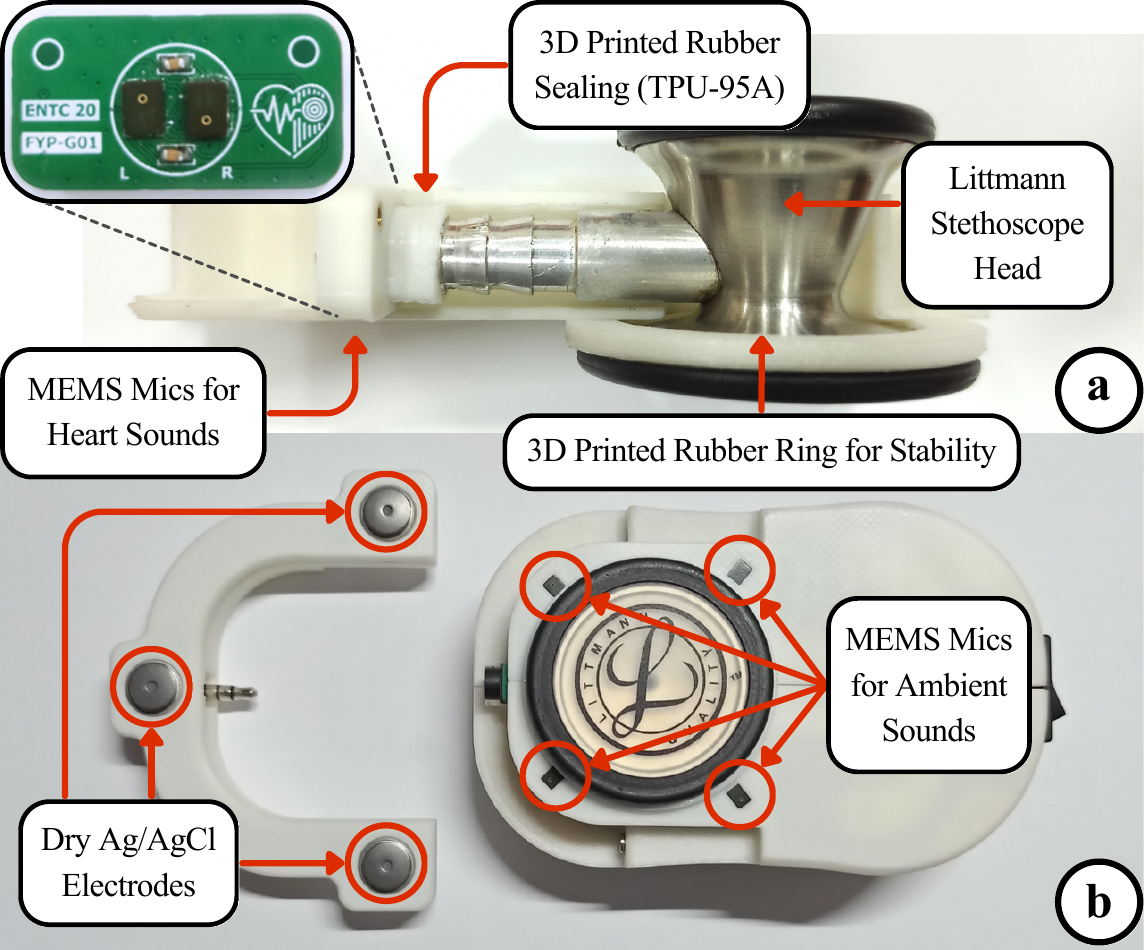}}
    \caption{(a) Cross-sectional view of the stethoscope-microphone interface for heart sound acquisition. (b) Bottom view of the device showing electrode interface and microphone placement for ambient sound acquisition.}
    \label{fig:prototype}
\end{figure}
\subsection{Simultaneous Acquisition of ECG and PCG Signals}

The device achieves its primary functionality of the main processing circuit (Fig.~\ref{fig:block}), using the following components:
a medical-grade ADS1294 IC~\cite{ads1294}, with a 24-bit $\Delta\Sigma$ ADC, is used for the single-lead ECG acquisition, while CMM-4030DT digital MEMS microphones~\cite{cmm-4030dt}, with a sensitivity of $-$26 dB and a bandwidth from 20 Hz to 1 kHz, are used for the heart and ambient sound acquisition. These mics generate a 1-bit PDM output, which is converted to a 20-bit value using an audio sample rate converter IC, ADAU7002~\cite{adau7002}. The STM32F746 microcontroller (MCU)~\cite{stm32f746} with an ARM Cortex M7 core is used for processing, while the RNBD451 Bluetooth\textregistered~Low Energy (BLE) module~\cite{rnbd451} is used for transmission of the processed data.

A 3M Littmann Cardiology III~\cite{littmann} stethoscope head is connected to a dual-microphone array to acquire PCG signals, with a rubber seal 3D-printed using TPU-95A and placed at the intersection to provide additional noise immunity. In this design, the stethoscope diaphragm provides acoustic amplification for the heart sounds before transduction, as these signals are too weak for a digital microphone to capture directly. A quad-microphone array is placed as shown in Fig.~\ref{fig:prototype}, with the sound ports facing outward, to effectively capture a reference of ambient noise while maintaining a compact device geometry.

\begin{figure}[!b]
    \vspace{-0.4cm}
    \centerline{\includegraphics[width=1\columnwidth]{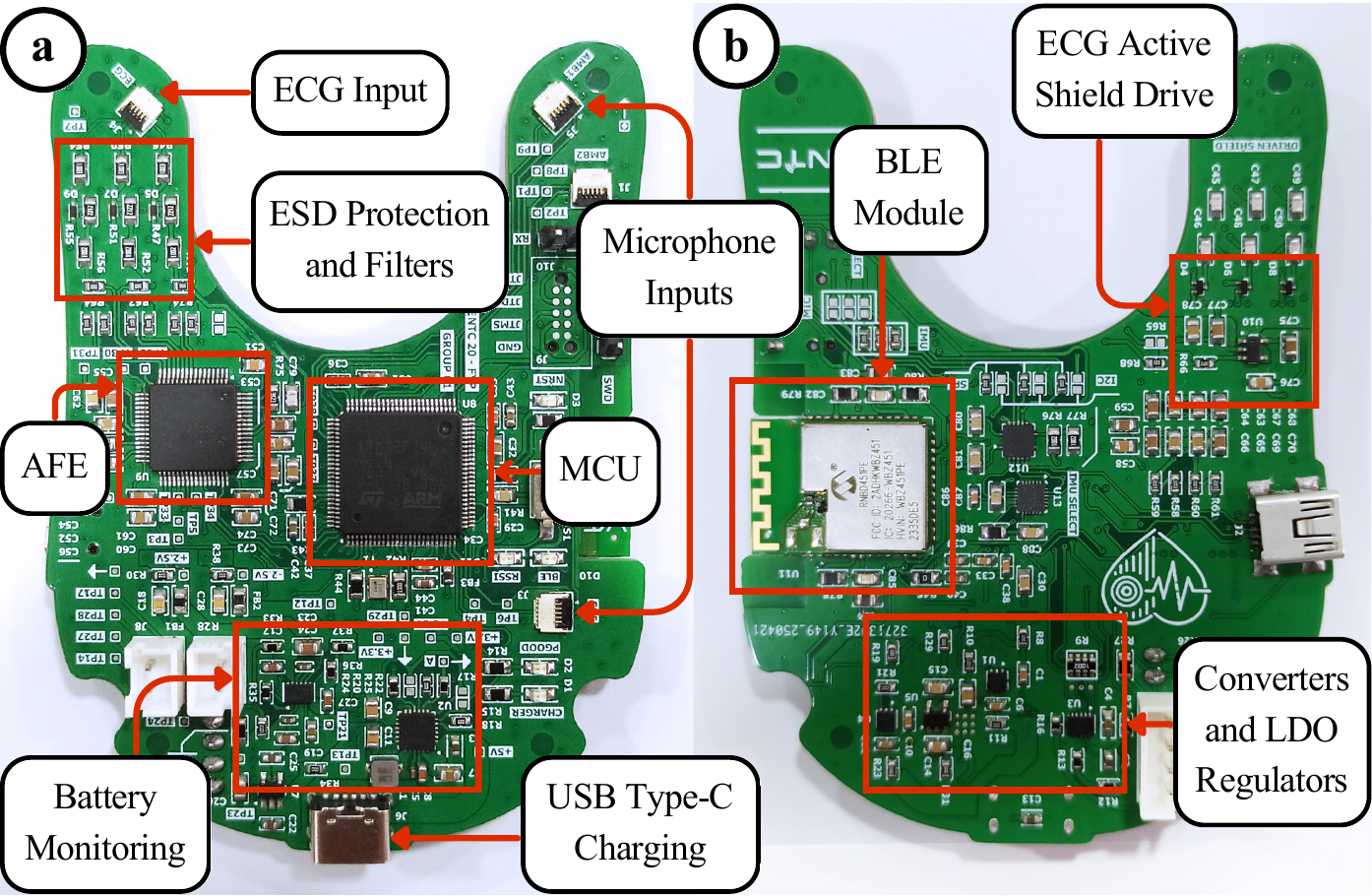}}
    \caption{(a) Top and (b) bottom views of the main controller board, highlighting the integration of the MCU, BLE module, and AFE, along with specialized circuitry for ECG sensing, acoustic signal acquisition, and power management.}
    \label{fig:pcb}
    \vspace{-0.4cm}
\end{figure}

The Lead-I ECG signal is captured using three Ag/AgCl dry electrodes, with one serving as the reference electrode to reduce common-mode noise. The placement of these electrodes is shown in Fig.~\ref{fig:prototype}. The electrode connection to the main device was designed as a detachable attachment, where a 3.5~mm jack connects this attachment to the main device, providing a low-noise pathway for the low-voltage analog signals to reach the analog front-end (AFE) IC. The MCU processes both ECG and PCG signals simultaneously, sampling ECG data at 500~Hz and the audio data at 8~kHz, which is subsequently downsampled to 2~kHz. The direct memory access feature is used to efficiently transfer audio samples from the peripheral to memory with minimal CPU intervention. This real-time data is then encapsulated into packets and streamed to the computer via the BLE module. The double-buffering technique is used to prevent data loss during acquisition, processing, and transmission. Fig.~\ref{fig:pcb} shows a labeled layout of the main controller board, including these subsystems.

\begin{figure}[!t]
    \centerline{\includegraphics[width=1\columnwidth]{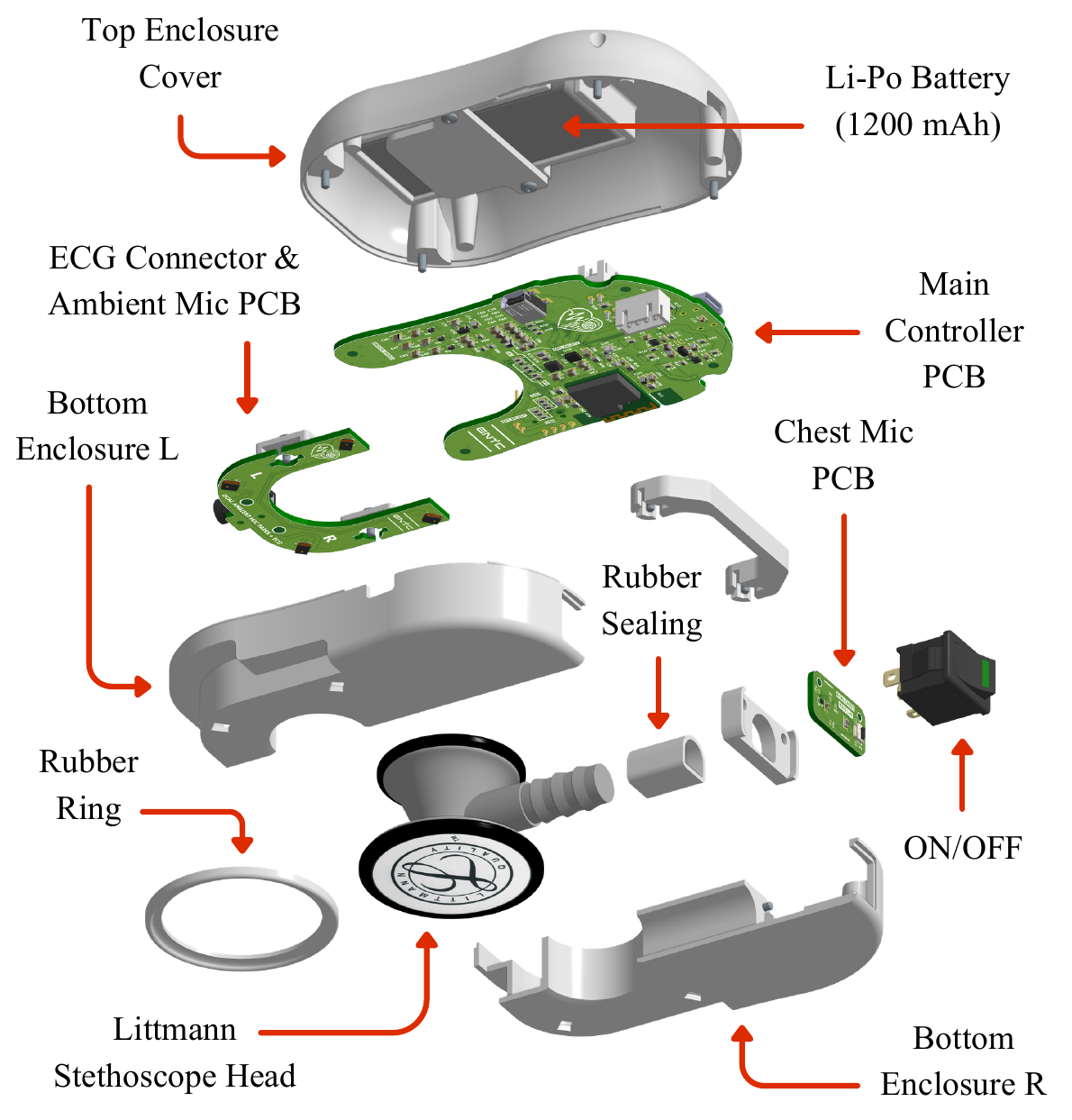}}
    \caption{Exploded CAD view of the prototype, illustrating the integration of the Littmann stethoscope head with the low-noise acoustic interface, ECG connectors, main controller board, and power management system.}
    \label{fig:exploded}
    \vspace{-0.4cm}
\end{figure}
The power management system is centered around a 3.7~V, 1200~mAh, 30~mm~$\times$~52~mm~$\times$~9~mm rechargeable lithium-polymer (\textit{Li-Po}) battery, managed by a custom-designed charging and monitoring circuit. System-wide power regulation is provided by $+$1.8~V and $+$3.3~V low-dropout (LDO) regulators, while a dedicated charge-pump-based bipolar supply generates $\pm$2.5~V specifically for the ECG acquisition circuitry. The realized device consumes a current of approximately 90~mA on average, with the runtime exceeding 13~h, despite not yet being optimized for low power consumption. An exploded view of the prototype computer-aided design (CAD) assembly is shown in Fig.~\ref{fig:exploded}. The device consists of two peripheral printed circuit boards (PCBs) for microphone placements and ECG connections. The dimensions of the complete prototype are 68~mm~$\times$~110~mm~$\times$~44~mm. making it easier to be used as a handheld device. The total fabrication cost of the PCBs, electronics, and 3D-printed components for the complete system prototype is approximately 190 USD.

\subsection{Real-Time Adaptive Noise Cancellation for PCG Signals}

A general LMS algorithm uses an adaptive finite impulse response (FIR) filter for noise cancellation~\cite{5264345}. In this implementation, the filter is used to denoise a recorded PCG signal, $x(n)$, using a reference signal, $r(n)$, which is uncorrelated with heart sounds but closely correlated with the noise.
To update the filter weights $\mathbf{w}(n)$, the LMS algorithm uses the method of steepest descent, given by
\begin{equation}
    \label{eq:w_update}
    \mathbf{w}(n+1) = \mathbf{w}(n) - \tfrac{1}{2} \mu \nabla e^2(n),
\end{equation}
where the step size $\mu$ controls the rate of convergence~\cite{5264345}.
As an improvement, the normalized LMS (NLMS) algorithm is used when the knowledge of input signal statistics is limited~\cite{236504}, which is often the case in real-time filtering. It uses a step size that is normalized by the reference input power and is given by:
\begin{equation}
    \mu = \frac{\mu_0}{\varepsilon + \|\mathbf{r}(n)\|^2}.
\end{equation}
Here, $\mu_0$ is the base step size, $\|\cdot\|^2$ is the Euclidean norm of the input vector, and $\varepsilon$ is a small positive constant.

However, in the conventional NLMS algorithm, sudden increases in the input noise power can cause the normalization term $\|\mathbf{r}(n)\|^2$ to become large, resulting in slower adaptation~\cite{article}.
To address this, we propose a BA-NLMS algorithm, which dynamically scales the step size during high-energy events.
In the real-time implementation, we estimate the average energy using a moving average expression given as:
\begin{equation}
    \bar{E}(n) = \alpha \bar{E}(n-1) + (1-\alpha)\|\mathbf{r}(n)\|^2,
\end{equation}
where $0 < \alpha < 1$ controls the averaging window.
The effective step size $\mu_{\text{eff}}(n)$ is then adapted dynamically based on the ratio between instantaneous and average energy:
\begin{equation}
    \mu_{\text{eff}}(n) =
    \begin{cases}
        \mu_0 \times \beta, & \text{if } \|\mathbf{r}(n)\|^2 > \eta \bar{E}(n), \\
        \mu_0, & \text{otherwise},
    \end{cases}
\end{equation}
where $\eta$ denotes the burst detection threshold and $\beta > 1$ is the step-size scaling factor.
Accordingly, by simplifying the gradient vector $\nabla e^2(n)$ into a product of $e(n)$ and $\mathbf{r}(n)$~\cite{5264345}, the weight update formula in \eqref{eq:w_update} can be expressed as:
\begin{equation}
    \mathbf{w}(n+1) = \mathbf{w}(n) + \left( \frac{\mu_{\text{eff}}(n)}{\varepsilon + \mathbf{r}^T(n) \cdot \mathbf{r}(n)} \right) e(n) \mathbf{r}(n).
\end{equation}

\begin{table*}[!t]
\centering
\caption{Performance comparison of proposed adaptive filtering method against existing real-time methods using NMSE, $\Delta$SNR, CC, and NMAE on a dataset of normal and abnormal PCG signals mixed with hospital noise at various SNR levels.}
\label{tab:compare_main}
\begin{threeparttable}
\begin{tblr}{
  cells = {c},
  cell{2}{1} = {r=3}{},
  cell{5}{1} = {r=3}{},
  cell{8}{1} = {r=1}{},
  cell{4}{3-6} = {bg=gray!20},
  cell{7}{3-6} = {bg=gray!20},
  vlines,
  hline{1,9} = {-}{0.08em},
  hline{2,5,8} = {-}{},
  hline{2,5,8} = {2}{-}{},
  hline{3-4,6-7} = {2-7}{},
}
\textbf{Signal Type} & \textbf{Filter Configuration} & \textbf{NMSE} $\downarrow$ & $\Delta$\textbf{SNR} (dB) $\uparrow$ & \textbf{CC} $\uparrow$ & \textbf{NMAE} $\downarrow$   \\
Normal               & LMS Filter~\cite{pauline2022robust} & 0.1576 $\pm$ 0.3527 & 15.4595 $\pm$ 4.8141 & 0.9456 $\pm$ 0.0798 & 0.2123 $\pm$ 0.2424 \\
                     & NLMS Filter                         & 0.0259 $\pm$ 0.0237 & 19.5678 $\pm$ 4.7743 & 0.9877 $\pm$ 0.0112 & 0.1304 $\pm$ 0.0487 \\
                     & Proposed BA-NLMS Filter             & 0.0229 $\pm$ 0.0201 & 20.0462 $\pm$ 4.9773 & 0.9891 $\pm$ 0.0099 & 0.1285 $\pm$ 0.0483 \\
Abnormal             & LMS Filter~\cite{pauline2022robust} & 0.0918 $\pm$ 0.2190 & 17.5187 $\pm$ 4.5253 & 0.9647 $\pm$ 0.0569 & 0.1541 $\pm$ 0.1782 \\
                     & NLMS Filter                         & 0.0285 $\pm$ 0.0238 & 19.2425 $\pm$ 5.2589 & 0.9863 $\pm$ 0.0119 & 0.1400 $\pm$ 0.0552 \\
                     & Proposed BA-NLMS Filter             & 0.0264 $\pm$ 0.0221 & 19.5504 $\pm$ 5.3933 & 0.9873 $\pm$ 0.0112 & 0.1386 $\pm$ 0.0546 \\
Mixed                & \textsuperscript{*}LU-Net (DL based) Framework~\cite{10173517} & 0.2864 $\pm$ 0.0829 & 5.5380 $\pm$ 1.5378 & NIA & NIA
\end{tblr}
\begin{tablenotes}
  \small
  \item \textsuperscript{*}Baseline results were evaluated on a different dataset compared to the rest of the study and are included for qualitative comparison.
  \item \hspace{0.13cm}NIA - No information available.
\end{tablenotes}
\end{threeparttable}
\vspace{-0.4cm}
\end{table*}

\begin{table*}[!b]
\vspace{-0.4cm}
\centering
\caption{Performance comparison of Burst-Adaptive NLMS against the NLMS algorithm using NMSE, $\Delta$SNR, CC, and NMAE on a dataset of normal and abnormal PCG signals mixed with burst-type noises at various SNR levels.}
\label{tab:compare_burst}
\begin{tblr}{
  cells = {c},
  cell{2}{1} = {r=2}{},
  cell{4}{1} = {r=2}{},
  cell{3}{3-6} = {bg=gray!20},
  cell{5}{3-6} = {bg=gray!20},
  vlines,
  hline{1,6} = {-}{0.08em},
  hline{2,4} = {-}{},
  hline{2,4} = {2}{-}{},
  hline{3,5} = {2-6}{},
}
\textbf{Signal Type} & \textbf{Filter Configuration} & \textbf{NMSE} $\downarrow$ & $\Delta$\textbf{SNR} (dB) $\uparrow$ & \textbf{CC} $\uparrow$ & \textbf{NMAE} $\downarrow$   \\
{Normal\\(Bursts only)}   & NLMS Filter              & 0.0546 $\pm$ 0.0487 & 17.3029 $\pm$ 3.2470 & 0.9747 $\pm$ 0.0207 & 0.1155 $\pm$ 0.0296 \\
                          & Proposed BA-NLMS Filter  & 0.0327 $\pm$ 0.0231 & 19.1962 $\pm$ 3.4049 & 0.9843 $\pm$ 0.0107 & 0.1043 $\pm$ 0.0279 \\
{Abnormal\\(Bursts only)} & NLMS Filter              & 0.0496 $\pm$ 0.0400 & 17.6148 $\pm$ 3.7049 & 0.9766 $\pm$ 0.0179 & 0.1160 $\pm$ 0.0349 \\
                          & Proposed BA-NLMS Filter  & 0.0316 $\pm$ 0.0194 & 19.2743 $\pm$ 3.7186 & 0.9848 $\pm$ 0.0093 & 0.1064 $\pm$ 0.0328 
\end{tblr}
\end{table*}

\subsection{Real-time Denoising Pipeline for ECG Signals}

The raw ECG signal is passed through three elliptic infinite-impulse-response filters: a 0.5~Hz high-pass filter to eliminate baseline drift, a 150~Hz low-pass filter to suppress high-frequency noise, and a 49.5--50.5~Hz bandstop filter to attenuate power-line interference. Elliptic filters were chosen for their sharp roll-off and high stopband attenuation at a low filter order, making them suitable for real-time processing.

\section{Experimental Results and Discussion}

This section describes dataset preparation and data preprocessing, hyperparameter sensitivity analysis, and comprehensive evaluation of the proposed algorithm through three validation steps as described below, ensuring robust performance across both controlled and clinical environments.
\begin{enumerate}
    \item \textbf{Comparative Performance Analysis:} Initial assessment using datasets of clean PCG signals augmented with real-world hospital noise, including a targeted analysis of performance during sudden noise bursts. Results are then compared against existing denoising methods.
    \item \textbf{Prototype Performance Validation:} Evaluations using the noisy clinical data recorded on patients at Lady Ridgeway Hospital (LRH) for Children, Colombo, Sri Lanka. This accounts for real-world acoustic coupling effects that artificial mixing may not fully capture.
    \item \textbf{Computational Complexity Analysis:} Assessing the algorithm’s complexity to ensure feasibility for real-time implementation on low-resource embedded platforms.
\end{enumerate}

\subsection{Dataset Preparation and Data Preprocessing}

\subsubsection{PhysioNet/CinC 2016}
This study uses the AUTHHSDB (training-\textit{c}) dataset of the publicly available PhysioNet/CinC Challenge 2016 database~\cite{Liu_2016} for clean PCG signals. It comprises 31 recordings sampled at 4~kHz with durations ranging from 9.6 to 122~s. From these, 14 clean recordings were randomly selected to create a balanced dataset comprising healthy subjects and patients with pathologies.

\subsubsection{Hospital Ambient Noise}
For the noise recordings, we use a hospital ambient noise dataset~\cite{10173517} comprising 562 real-world recordings sampled at $44.1$~kHz, recorded from different areas of a busy hospital (corridors, waiting rooms, etc.).

\subsubsection{Synthetic Dataset Preparation}
Both clean PCG and noise recordings were normalized and downsampled to 2~kHz to simulate our device behavior. The noise signals in the dataset are each 5~s long. Therefore, the noise is matched in length to the clean signals by repetition. This recording is passed through an autoregressive filter to create a noise signal and, separately, through a moving-average FIR filter to create a reference signal~\cite {mathworksLMS}. The noise signal was then scaled and mixed with clean PCG recordings at random SNR levels from $-$10 to 5~dB to simulate the mixture signal, $x(n)$. This effectively creates a noisy dataset of 3934 recordings, each for normal and abnormal heart sounds, which are then used to evaluate the proposed filters' denoising capabilities.

\subsubsection{Clinical Dataset Preparation}
\begin{figure}[!b]
    \vspace{-0.4cm}
    \centerline{\includegraphics[width=1\columnwidth]{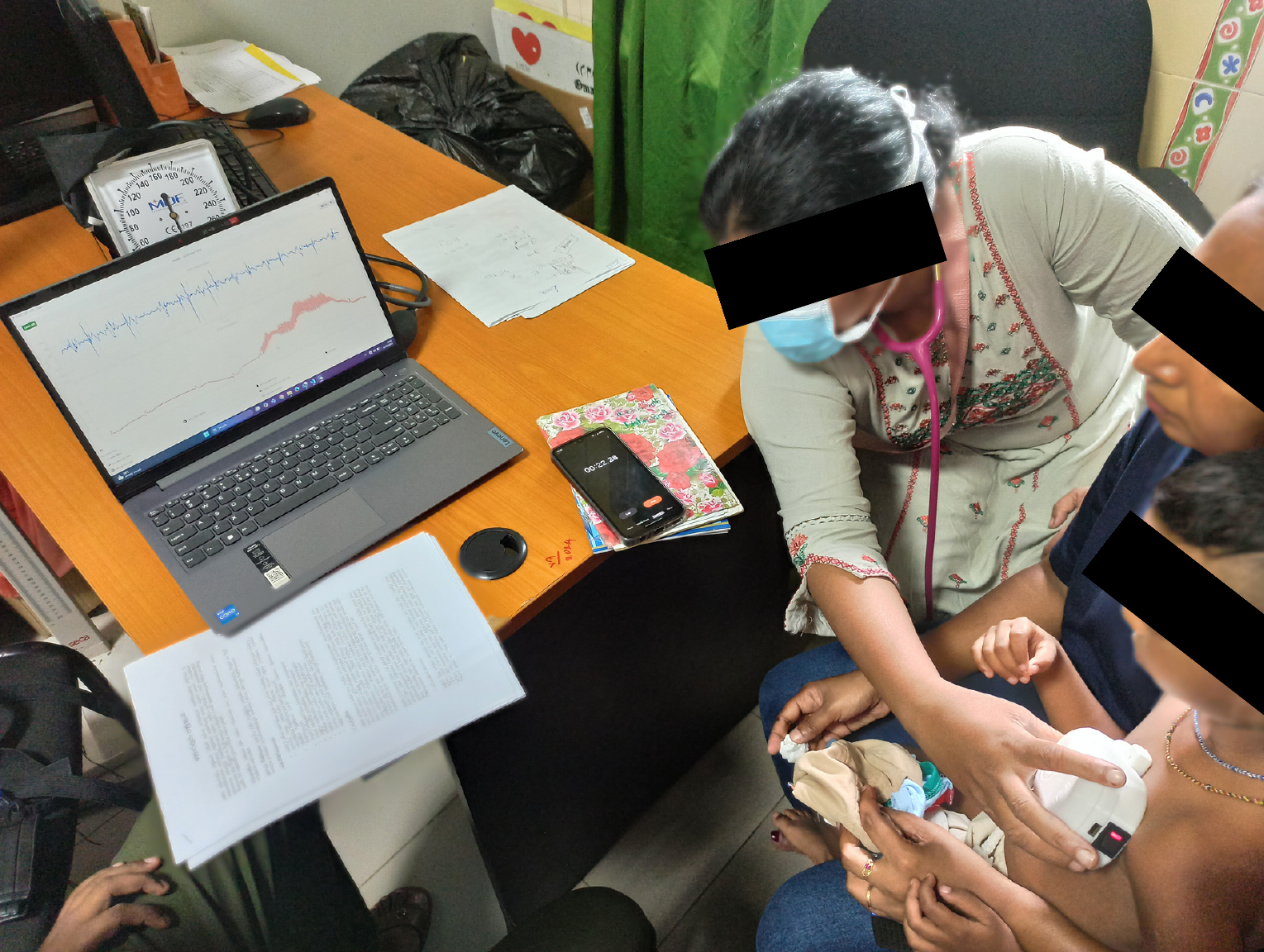}}
    \caption{Clinical validation of the proposed cardiac screening device conducted at LRH, Colombo. The study involved simultaneous ECG and PCG data acquisition to evaluate the performance in a typical noisy hospital setting.}
    \label{fig:lrh}
\end{figure}

We recorded simultaneous ECG-PCG data in a noisy hospital environment at LRH, Colombo, using our dual-modality device. The data collection process was standardized in accordance with the “University of Moratuwa Ethics Review Committee Regulations” (Ethics Declaration Number: ERN/2025/002) and the “Lady Ridgeway Hospital Ethics Review Committee Regulations” (App. Number: LRH/DS/29/2022). The collected data were used as research samples, and all subjects were informed and provided consent. A photograph obtained during the study at LRH is shown in Fig.~\ref{fig:lrh}, where data were recorded in a hospital setting.

\subsection{Hyperparameter Sensitivity Analysis}

The hyperparameters $\alpha$, $\eta$, and $\beta$ must be tuned based on the characteristics of the noisy environment.

\subsubsection{Smoothing Factor (\texorpdfstring{$\alpha$}{alpha})}
Typically chosen close to 1 to maintain a stable baseline. We note that a higher $\alpha$ ensures the average energy estimate does not react to bursts, thereby increasing the sensitivity of burst detection. However, if $\alpha$ is too high, the algorithm becomes extremely slow in tracking actual changes in the average energy. Conversely, a lower $\alpha$ would reduce the sensitivity of burst detection, as the threshold changes continuously with the changes in average energy.

\subsubsection{Burst detection threshold (\texorpdfstring{$\eta$}{eta})}
This parameter controls the trigger for burst adaptation. A lower $\eta$ makes the system overly sensitive even to slight random fluctuations, which can hinder the convergence of the NLMS algorithm~\cite{a15040111}. A higher $\eta$ makes the algorithm less sensitive to burst noise, leading to slower convergence as the system remains in the standard NLMS mode.

\subsubsection{Step-size scaling factor (\texorpdfstring{$\beta$}{beta})}
Typically selected to be slightly greater than 1 to facilitate burst adaptation. However, excessively high values can result in filter instability and high steady-state error. To satisfy convergence criteria defined in~\cite{236504}, $\beta$ must be constrained such that $\mu_0 \beta <$~2.

In this study, the filter is evaluated with a length of 10, and parameters: $\mu_0=$~0.05, $\alpha=$~0.99, $\eta=$~5, $\beta=$~6, and $\varepsilon=$~0.001. The parameters were empirically chosen based on the trial-and-error method. For comparison, the standard LMS and NLMS filters were also evaluated with the same parameters.

\begin{figure}[!t]
    \centerline{\includegraphics[width=1\columnwidth]{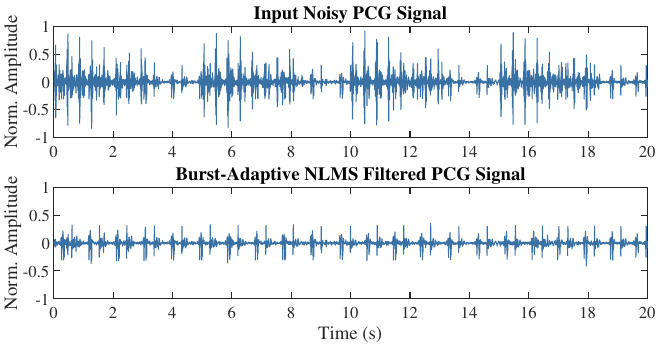}}
    \caption{Denoising performance of the Burst-Adaptive NLMS filter on a PCG signal. A clean signal was corrupted by burst noise to an input SNR of -3~dB. The filtered output shows an $\Delta$SNR of +19.31~dB and a CC of 0.9889.}
    \label{fig:burst_noises_TD}
    \vspace{-0.4cm}
\end{figure}
\label{sec:dataset}
\subsection{Comparative Analysis with Artificially Induced Noise}

Four metrics, which are independent of the input signal power, were used for evaluation: normalized mean squared error (NMSE), signal-to-noise ratio improvement ($\Delta$SNR), correlation coefficient (CC), and normalized mean absolute error (NMAE).
As shown in Table~\ref{tab:compare_main}, the BA-NLMS algorithm clearly outperforms both standard algorithms for all normal and abnormal PCG recordings. Also, when analysed qualitatively, it has a significant improvement over the real-time DL based approach as well. Even though recently proposed methods such as the robust multistage LMS filter~\cite{pauline2022robust} and greater cane rat algorithm (GCRA)-optimized ANC~\cite{alla2025enhancing} may achieve better denoising performance, they are not appropriate for real-time implementations.

To specifically evaluate performance against burst-type noise, 9 recordings containing sudden bursts were selected from our synthetic dataset and were assessed. As summarized in Table~\ref{tab:compare_burst} and illustrated in Fig.~\ref{fig:burst_noises_TD}, the BA-NLMS filter demonstrated a clear improvement over the standard NLMS filter. This is critical for handling non-stationary hospital noise, which is characterized by sudden amplitude changes.

\subsection{Performance Validation with Noisy Recorded Data}

\begin{figure*}[!t]
    \centerline{\includegraphics[width=1\textwidth]{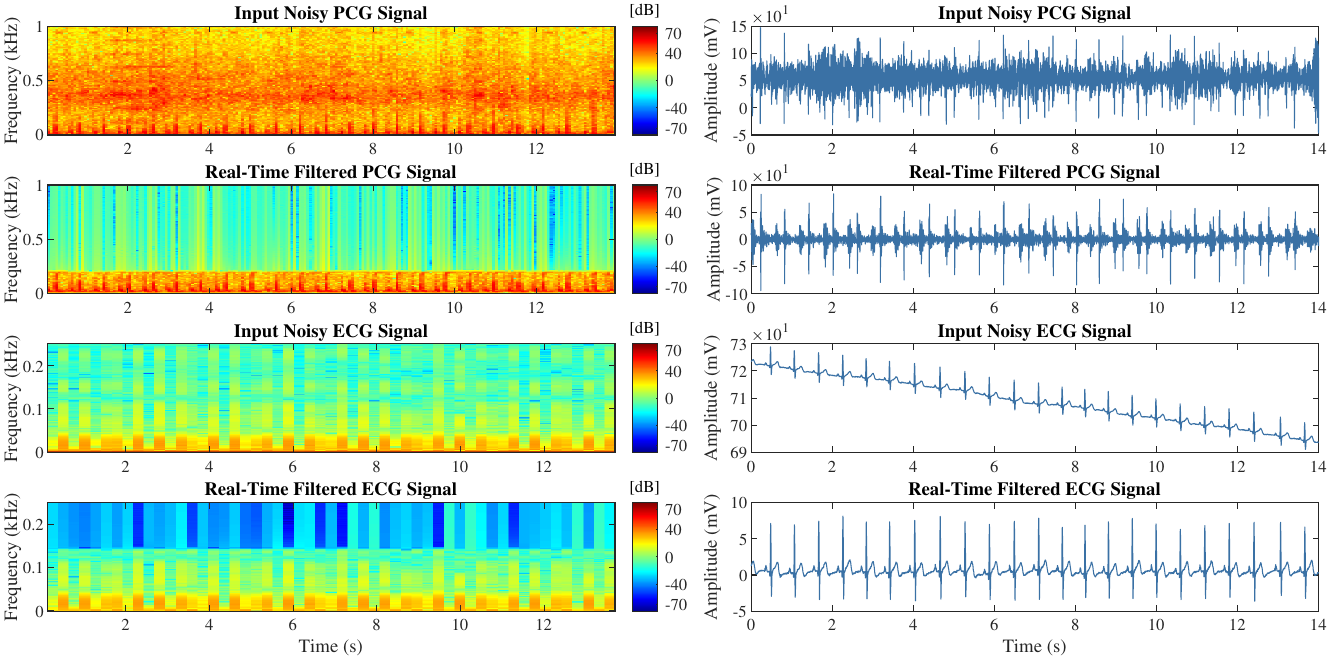}}
    \caption{Denoising performance of the proposed real-time filter algorithms on recorded noisy ECG and PCG signals. Spectrograms (left) and time-domain plots (right) show the improvement in the noisy ECG (input SNR of -9.08~dB) and noisy PCG (input SNR of -2.08~dB) signals, respectively.}
    \label{fig:filt_ECG_PCG}
    \vspace{-0.4cm}
\end{figure*}
To test real-world performance, we used the clinical data and then adaptively denoised it. The last stage of our PCG denoising pipeline includes a 20--200~Hz band-pass filter, targeting the primary heart sound range~\cite{zang2025novel}.
Fig.~\ref{fig:filt_ECG_PCG} demonstrates the spectrograms and the time domain plots of the denoising outputs of both PCG and ECG signals, respectively.
We evaluated the performance of our prototype by using simultaneous signal segments having a fixed time period of 14~s each, and calculating the SNR using:
\begin{equation} 
\textit{SNR} = 10\,\cdot \log_{10}\left[\frac{\textit{power}~{(LF)}}{\textit{power}~{(LFN)}+\textit{power}~{(HF)}}\right],
\end{equation}
where a low frequency range ($LF$) is expected to contain useful data, whereas a low frequency noise ($LFN$) and a high frequency range ($HF$) are expected to contain noise. The BA-NLMS filter pipeline improved the noisy PCG signals by an average $\Delta\text{SNR}$ of 37.01~dB while the noisy ECG signals were also improved, achieving an average $\Delta\text{SNR}$ of 30.32~dB.

\subsection{Algorithm Complexity Analysis for Embedded Feasibility}
The computational complexity of the proposed BA-NLMS algorithm is analyzed with respect to the filter length $L$.

\subsubsection{Time Complexity}
For each incoming sample, the algorithm performs buffer shifting, two dot-product operations, and a weight vector update. Each of these steps requires $\mathcal{O}(L)$ arithmetic operations. The auxiliary operations, including burst detection and step-size selection, involve only scalar computations and thus maintain a constant $\mathcal{O}(1)$ time complexity. Consequently, the proposed algorithm maintains a per-iteration time complexity of $\mathcal{O}(L)$, resulting in a total computational complexity of $\mathcal{O}(N \cdot L)$ for a sequence of $N$ input samples.

\subsubsection{Space Complexity}
The space complexity is linear relative to the filter length $L$. The memory footprint is $\mathcal{O}(L)$, primarily determined by the persistent storage required for the filter weights and the input buffer.

Since the proposed BA-NLMS algorithm preserves the $\mathcal{O}(L)$ asymptotic computational and memory complexity of the conventional NLMS algorithm, it introduces no significant computational overhead. Furthermore, the per-sample processing time scales linearly with the $L$ and remains constant with respect to the input signal duration. Hence, the algorithm is well-suited for real-time embedded implementation on resource-constrained platforms, such as MCUs and low-power digital signal processors.

\section{Conclusion and Future Works}

We have proposed a portable device capable of simultaneous ECG and PCG acquisition with integrated real-time ANC. The system incorporates a novel real-time BA-NLMS algorithm that effectively handles non-stationary, burst-type noise, achieving an average SNR improvement of 37.01~dB, while the real-time ECG filter achieved an average SNR improvement of 30.32~dB on a recorded clinical dataset, demonstrating the robustness of the proposed system, while also maintaining a lower complexity for potential embedded implementation. By providing a comprehensive, synchronized view of the heart's contractile function, the device empowers healthcare workers to immediately and reliably identify subtle murmurs, significantly enhancing diagnostic accuracy compared to traditional auscultation alone. Future works will focus on (a) further improving the SNR performance of the prototype device while evaluating in various hospital environments, (b) implementing the denoising algorithms on the microcontroller for fully embedded operation, and (c) developing an ECG-PCG dataset recorded using the proposed device to support future machine learning-based diagnostic studies. Ultimately, this system has the potential to be a valuable tool for enabling more reliable cardiac assessments in both clinical and low-resource environments, and expanding affordable community screening for pre-existing heart diseases.

\section{Acknowledgements}

The authors thank the staff of Lady Ridgeway Hospital for Children for providing access to record data for prototype performance evaluation and Voex Technologies Pvt. Ltd. for funding the prototype development.

\bibliographystyle{IEEEbib}
\bibliography{refs}

\end{document}